# On the Roles of the Secondary Circulation in the Formation of Hurricanes

Chanh Q. Kieu[+]
*Department of Atmospheric and Oceanic Science, University of Maryland, College Park*

**Abstract**

Secondary circulations (SC) associated with hurricanes are traditionally regarded as small perturbations superimposed on the primary circulations (PC). The reason behind this treatment roots in an observation that the magnitude of the SC is about 10 orders of magnitude smaller than that of the PC. This serves a wide range of hurricane theories up until now with some considerable successes. Recently, Kieu (2004) proposes a revitalizing theory for the development of hurricanes in which the author is able to obtain a class of exact solutions of the primitive equations which shares some of the most important dynamical aspects with observations. According to this theory, the SC turns out to be particular important in determining the three-dimensional structure and time evolution of axisymmetric hurricanes. Kieu's theory, however, contains an infinite growth of the SC that all hurricane theories so far encounter in attempt to modeling the development of hurricanes. In this study, it will be shown that the infinite growth does not occur. In fact, the solution becomes stationary after a period of time and the SC is able to maintain itself without blowing exponentially with time if the nonlinear terms in the vertical momentum equation are included. In addition, the SC tends to force the peripheral convection to converge toward the center, and thus builds up a more concentric vortex with a typical hurricane eye structure. Some potential roles of SC in the formation of hurricane eyes are also discussed.

## 1. Introduction

The contemporary approach to hurricanes insofar still considers the secondary circulation (SC) associated with the radial and vertical motions as a second-order approximation superimposed on the primary circulation (PC). The main reason for this approximation is due to the weakness of the SC compared with the PC (about 10 orders of magnitude smaller). Typical examples are the works of Charney and Eliassen (1964), Yanai (1964), Willoughby (1979), Montgomery and Farrell (1993). Based on this scale analysis, one is able to expand the primitive equations in terms of some dimensionless numbers, for example the ratio of radial wind over tangential wind. The SC is then treated as an isolated system with the PC acting as a known basic flow. This treatment underlies almost all of the hurricane theories up to now with some considerable successes, especially during the mature stage of a hurricane. The immediate limitation in this approach is clear: *it does not allow one to understand how the SC interacts with the evolution of the PC*. At least to date, no such detailed understanding is known. In this sense, the SC is just a passive agent in the whole development of hurricanes.

Recently, Kieu (2004, hereafter referred to as K04) presents a refreshing theory

---

[+] *Corresponding author address*: Chanh Q. Kieu, Department of Atmospheric and Oceanic Science, University of Maryland, College Park, MD 20740. Email: kieucq@atmos.umd.edu

for the hurricane development which is completely different from the previous approaches. Instead of using the classical "perturbation" method, K04 tackles directly the system of primitive equations with a simple assumption of positive feedback between latent heat release and vertical motion. With some simplifications, a class of analytical solutions for both SC and PC during the intensification period of axisymmetric vortices is obtained exclusively, which share some striking similarities with observations. In this theory, the SC turns out to be of notable importance. In fact, it is the vertical advection by the SC that decides the whole vertical structure of the PC as well as the different intensification of the PC at the small and large radius limits. The contributions from the SC to the intensification of the PC inside the radius of maximum wind (RMW) are so important that neglect the SC will result in a PC that is no longer comparable with the observations (Willoughby 1982; McBride 1981a, b; K04).

One of the most important observations in K04's theory is that given a feedback mechanism relating the heating rate with vertical/radial motions, one is able to obtain the SC entirely independent from the PC. This is because the system of the primitive equations is separated into two subsystems; one involves the vertical momentum equation (VME) and the continuity equation, the other consists of radial and tangential momentum equations. Given the solution for the SC from former subsystem, the development and structure of the PC will be followed subsequently. The assumption of the feedback is a key thing in the whole argument because it assumes that the heating rate is a function of vertical motion only. More latent heat release will result in stronger upward motion, which in turn can induce more latent heat released. In reality, the feedback mechanism roots in the moisture convergence flux and this flux is related closely to the PC at the surface (Emanuel 1986). By assuming the feedback mechanism, the connection between moisture flux and the PC is automatically taken into account (kind of parameterization). However, this connection should not be taken too far. K04's model employs no Reynolds averaging operator. Therefore, no eddy terms in any way can appear and it is therefore unnecessary to have a truly parameterization for the eddy terms in all equations.

Several points should be pointed out in K04's theory. The first is that, like all previous theories for the development of incipient storms (e.g. Charney and Eliassen 1964; Yanai 1964), the SC associated with the storm grows infinitely as an exponential function of time. Although the theory may apply for a short period of time for which the exponential grow may be expanded as a linear function, the growth will eventually blow up and the theory is no longer applicable. Second, the nonlinear terms (NLT) in the vertical motion have been neglected to make the system of the primitive equation solvable. This is an unexpected feature of the theory and, as noted in K04, it is closely related to the infinite growth of the SC. The insurmountable complications when all NLTs in the VME are included make the system of nonlinear equations unsolvable and K04 has to content with the limitations of the theoretical solutions.

In this work, the full nonlinear VME will be employed to study the behaviors of the SC with time under the same positive feedback mechanism used in K04. The price we have to pay for using the full VME is that we now have to appeal to numerical methods to understand the behaviors of solutions. The inclusion of the NLTs in the VME turns out to be especially vital because this results in a stationary solution for the SC, a kink of the VME, after some period of time instead of evolving infinitely. The implication of this result is significant. In fact, it supports the K04's theory in the sense that the SC cannot be merely regarded as a simple perturbation superimposed on the PC as in previous studies. If one concerns with the three dimensional structure as well as the time evolution of hurricanes, the SC has to be taken fully into account. Of interest, the stable solution of the SC shows a striking connection with the formation of hurricane eyes. Furthermore, the SC confirms the convergence of the peripheral convection from outside toward the center, and this builds up a concentric vortex with typical hurricane structures. It is worth to note promptly at this point that any numerical method, one way or another, contains some weaknesses and/or spurious solutions. To eliminate as many weaknesses of a numerical method as possible, different sensitivity experiments are performed to ensure that the solutions obtained are physically good. In this sense, this work is not a complete mathematical proof of the existence of the stationary solution of the SC, but instead provides a numerical evidence that the solution does appear.

A small convention in this work: the author will refer generally an axisymmetric vortex as a hurricane for the ease of argument. The main focus here is a transition from tropical depression to hurricane stage and it is more appropriate to use the term "tropical storm" to describe the developing vortex. However, no such detail is made here for simplicity.

The structure of this paper is as follows. The basics equations used in this work are given in section 2. Model setups and numerical scheme as well as boundary and initial condition are addressed in section 3. Section 4 presents model results and their implications. Section 5 provides several sensitivity experiments. The implicit roles of the SC on the formation of hurricane eyes are discussed in section 6, and the conclusions are given in the final section.

2. **Basic equations**

The basic equations in this study are an extended version of the ones used in K04 with all the NLTs in the VME included (these NLTs are neglected in K04). Because the main purpose is to demonstrate the structure and temporal evolution of the SC, the full system of primitive equation will not be necessary. This is because the VME and the continuity equation comprise a close system under the feedback assumption between latent heat release and vertical motion, and they are enough to resolve completely the SC. The PC will follow by using the same procedures as in K04 and it will not be addressed here. The two equations employed in this study are:

$$\frac{\partial w}{\partial t} + u\frac{\partial w}{\partial r} + w\frac{\partial w}{\partial z} = J - Nw \qquad (1)$$

$$\frac{1}{r}\frac{\partial (ru)}{\partial r} + \frac{\partial w}{\partial z} = 0 \qquad (2)$$

where J is a source term representing vertical forcings (the source term J contains all possible contributions such as buoyancy force, perturbation pressure gradient, etc.), N is proportional to the Brunt-Väisälä frequency, w and u are vertical and radial winds, respectively. Except for the assumption of incompressibility of the atmosphere, these two equations are exact because no approximation was made. There is no azimuthal dependence in Eqs. (1) and (2) because hurricanes are assumed to be axisymmetric[1]. As long as the feedback process (the J term) is related solely to vertical motion (or radial motion), these two equations are closed: given the vertical wind w, the radial wind u will follow immediately and vice versa. The feedback can be expressed by: J = kw, where k is a positive constant coefficient, and it is expected to be effective when vertical motion is greater than some threshold value. Noted that both Eqs. (1) and (2) are not subject to any averaging operator. Therefore, the variables are exact and no eddy terms appear in the equations. More discussion can be found in K04.

As mentioned by Smith (1980), the source term J on the RHS of Eq. (1) consists of two dominant terms: one is the buoyancy force and the other is usually referred to as the perturbation pressure. Because these two terms are nearly of the same order of magnitude (Zhang et al. 2000), it is supposed to be inaccurate to integrate Eq. (1) directly with time, and vertical wind is often calculated diagnostically by using the continuity equation. In this study, there is, however, no such separation between the buoyancy forcing and perturbation pressure. The source term J on the RHS of Eq. (1) represents the total forcing.

3. **Model setup**

*Numerical schemes*

It is a huge challenge to find an exact solution of Eqs. (1) and (2). Numerical methods are therefore the most useful way to obtain some behaviors of the solution of these equations. In this study, a 2D simulation in the radius-height cross section is applied. The Runge-Kutta 4$^{th}$ order scheme will be used to integrate Eqs. (1) and (2) for up to 12 hours. The model consists of 2001 x 71 grid points in r and z dimensions with dr = 500 m and dz = 300 m. The feedback between J and w is turned on if vertical motion is greater than some threshold value, which is set to equal to 1 ms$^{-1}$ in this study. Various model parameters are given in table 1. These default values are considered as a control run. Several supplementary sensitivity experiments will be addressed in section 5.

---

[1] According to the author point of view, the axisymmetry of hurricanes should be considered as an internal feature of hurricanes instead of as an assumption. All asymmetric features are due to the inhomogeneous environment that the hurricanes are embedded in.

One of the difficulties in solving Eqs. (1) and (2) by numerical methods is the generation of internal gravity waves due to the stable stratification of the atmosphere. These waves propagate vertically and horizontally, carrying significant amount of energy of the system out of the model domain. Numerically, these waves are the source for the instability of the model and cause the model to overflow after few time steps. To overcome this difficulty, a high-frequency wave filter is applied to get rid of the waves. In addition, sponge layers are added at the top and R = 1000 km to represent the radiation of the waves into outer space and to prevent the reflections of the waves from the boundaries. Even though this filter is a numerical process introduced to stabilize the integration, it is worth to emphasize that, to some extent, the filter is very necessary to account for the actual loss of the energy in the real atmosphere consumed by the waves.

*Initial condition*

The initial condition for the vertical motion is assumed of the following form:

$$w(r,z,0) = w_0 e^{-(r-a)^2/R_0^2} \sin(\frac{\pi z}{H}) - \varepsilon \qquad (3)$$

where $R_0$ is a scale radius beyond which the vertical motion is substantially small, $w_o$ is the scale of vertical motion, H is the depth of the troposphere, $\varepsilon$ is a small positive constant accounting for the descending motion at very large radius. Physically, this $\varepsilon$ guarantees the conservation of mass inside a closed domain: a very strong upward motion within a small area near the center of the storm is compensated by a very weak but broad descending motion at the large radius. This descending motion also appeared in previous studies, e.g., as one of the cycle legs in Emanuel's theory (1986) or as an initial condition in hurricane simulations. The sine function in Eq. (3) takes into account the impenetrability of the surface and the upper tropopause. A justification for using the functional form (3) is that during the genesis stage of a hurricane, there usually appears a Mesoscale Convective Systems (MCS, see e.g., Zhang and Fritch 1986) with a broad cloud cluster hovering around. The cloudy area signals substantially stronger convective activities inside than those outside the cluster. The function (3) satisfies this observation fairly well. Different initializations sharing the same structure with (3) have been tested and the differences in the results are minor. Experiment 3 in Section 5 provides an example in which an ensemble of small-scale cumulonimbus columns is utilized to initialize the model. Having obtained the vertical wind, the radial wind will follow by simply integrating the continuity equation outward with respect to the radius. For the stratification of the atmosphere, a simple function of the form $N = \chi \exp(-\mu^2)$ is selected, where $\chi$ is a constant and $\mu = (H_0-z)/Z_0$ ($H_0$ is the height of the top of the model, and $Z_0$ is a scale of stratification). The values of these parameters are provided in table 1. This gives a strong stable upper atmosphere but virtually neutral at the lower troposphere. A sensitivity experiment for the stratification is also provided in Section 5.

*Boundary condition*

Consider first the radial wind at the left boundary, i.e. $r = 0$, and at the outer boundary, i.e. $r = R$ where R is the farthermost radius. Apparently, u must be equal to zero at $r = 0$ due to the axisymmetric characteristic of hurricanes. However, it is not required to have the boundary condition for the radial wind at $r = R$ because the continuity equation is of the first order. One just simply needs to integrate the continuity equation outward and the radial wind at $r = R$ is obtained automatically. For the bottom and top boundaries, it is reasonable to utilize the Neumann boundary (NB) condition for the radial wind:

$$u_{z=0} = \alpha_b u_{z=dz} \quad \text{and} \quad u_{z=H_0} = \alpha_t u_{z=H_0-dz} \tag{4}$$

where $\alpha_b$ and $\alpha_t$ are proportional constants. This boundary is consistent with K04's theory for which no frictional effects are included (free-slip boundary). The bottom and top boundary conditions for the vertical wind are set equal to zero (rigid boundary condition). At the outer boundary $r = R$, w is supposed to have a NB type (i.e, $w|_{r=R} = w|_{r=R-dr}$). The boundary condition for w at the center $r = 0$, nevertheless, should be paid particular attention. Using Eq. (1) and the boundary condition for the radial wind at $r = 0$, one obtains an equation for w at $r = 0$, which no longer contains the radial advection term. Eq. (1) now becomes a 1D hyperbolic equation that has no physical contact with the surrounding environment (unless the diffusivity is taken into account at $r = 0$). The vertical motion at $r = 0$ thus propagates just up and down between the top and bottom boundaries. This happens because we have here a totally axisymmetric model and the kinematical point $r = 0$ turns out to be completely isolated. The most reasonable assumption here for the vertical motion at the point $r = 0$ is $w_{r=0} = \eta w_{r=dr}$, where $\eta$ is a proportional constant. In the case of ideal fluid, $\eta$ is equal to zero and this value will be the default one in the control run. A sensitivity experiment with different $\eta$ is given in Section 5 (experiment EXP 2).

**Table 1**

| Parameter | Description | Value | Unit |
|---|---|---|---|
| $U_0$ | Scale radial wind | 10 | $ms^{-1}$ |
| $w_0$ | Scale vertical wind | 8 | $ms^{-1}$ |
| H | Depth of the troposphere (cf. Eq. 3) | 15 | km |
| $H_0$ | Model height (used in the stratification distribution) | 21 | km |
| dr | increment in radius | 500 | m |
| dz | increment in height | 300 | m |
| dt | time step | 5 | s |
| R | Model dimension in r-direction | 1000 | km |
| $R_0$ | Scale radius of initial cloud cluster (cf. Eq. 3) | 50 | km |
| k | feedback coefficient (used in the feedback relationship) | $10*U_0/R_0$ | $s^{-1}$ |
| Rtime | forecast time | 12 | hour |
| $w_{thres}$ | threshold at which feedback is effective | 1 | $ms^{-1}$ |
| $\varepsilon$ | Compensating vertical motion (cf. Eq. 3) | 0.001 | $ms^{-1}$ |
| $\alpha_b$, $\alpha_t$ | Coefficients for the bottom and top boundaries (cf. Eq. 4) | 1 | |
| $M_r$ | Number of spongy layers in r-direction | 4 | |
| $M_z$ | Number of spongy layers in z-direction | 4 | |
| $D_r$ | Diffusive coefficient within the spongy layer (r-direction) | $0.1*dr^2/dt$ | $ms^{-2}$ |
| $D_z$ | Diffusive coefficient within the spongy layer (z-direction) | $0.1*dz^2/dt$ | $ms^{-2}$ |
| $\chi$ | Scaling stratification of the atmosphere | $10^{-3}$ | $s^{-1}$ |
| $Z_0$ | Scale of the vertical distribution of the stratification | 5000 | m |
| a | cf. Eq. (3) | 10 | km |
| Nx | Number of grid points in r-direction | 2001 | |
| Ny | Number of grid points in z-direction | 71 | |
| f | Coriolis parameter | $10^{-4}$ | $s^{-1}$ |

4. **Results**

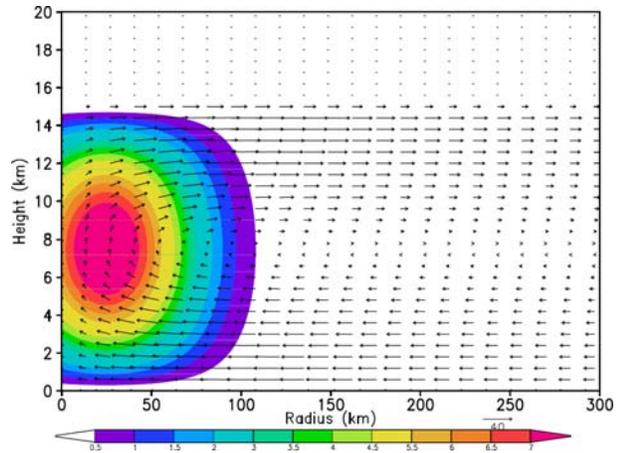

Figure 1. Radius-height cross section of the vertical motion (shaded) superimposed by the in-plane flows at the initial time t = 0.

Fig. 1 shows the initial condition for both vertical and radial winds using the profile (3). This configuration corresponds roughly to the early stage of a hurricane with a broad cloud cluster as well as a fairly weak vertical motion at the center, which can be found in, e.g. Palmen and Newton (1969). This initialization results in a large radial wind up to 40 $ms^{-1}$ at the surface because the ascending motion occurs *everywhere* within the incipient cluster according to the profile of vertical motion given by expression (3). A more realistic initialization with an ensemble of VHTs (experiment EXP 3 in Section 5) will give more reasonable radial flow.

Fig. 2 shows a time series of the vertical motion after the model is integrated. The SC accelerates rapidly for the first 30 minutes under the contributions from both the feedback mechanism and the adjustment of the initial unbalanced flow (the adjustment is unavoidable because the vertical motion given by (3) and the radial wind obtained by integrating the continuity equation do not satisfy the VME initially. Therefore, the adjustment is a must). However, the growth cannot intensify infinitely because the NLTs in the VME will promptly start their roles. First, the vertical/radial advections will adjust the SC so that the circulation approaches a more stable configuration. The unbalanced flow quickly dissipate (e.g. Gill 1982), leaving behind the more stable structure as seen in Fig 2b. Secondly, the NLTs excite internal gravity waves that propagate throughout the domain. A considerable amount of the total energy is converted into the wave activities, and these waves are supposed to radiate away in the real atmosphere, consuming the energy of the system quickly. In the model, this wave-related energy sink is taken into account by the filtering operators. The role of the filter is thus dual. On one hand, it is introduced to eliminate the high-frequency waves and thus makes the model integrable. On the other hand, it reflects a real process in which gravity waves propagate away from the system, consuming significantly the total energy. After 5-h integration, the SC attains a stable structure for which the energy from the latent heat release is balanced by the energy carried away by the gravity waves. No infinite growth occurs. Even though the radial wind is fairly large at the initial time, the stable SC at the later time shows a reasonable magnitude and structure. The stability of the atmosphere forces the SC to confine within the troposphere, which bears strong resemblance with observations. Of interest, the eyewall structure is built up immediately after the model is started (cf. Figs. 1 and 2a) and the pattern keeps surviving continuously at later time. The evolution seen in Fig. 2 reveals one important feature: peripheral convection is forced to converge inward, creating a typical hurricane-eye structure. This is significant because this tells us that the SC itself has capability to gather convections from outside to feed its development. The eyewall structure, as will be shown in section 5, is however merely due to the boundary condition imposed on w at r = 0. It is hard for anyone to imagine a kinematical boundary at one mathematical point r = 0 can decide the structure of hurricane eyes. However, this result may have more implications than it appears because the formation of hurricane eyes may in fact depend much on whether an incipient cloud cluster is able to find an area with essentially no vertical motion to build the eye around. Those calm areas are not rare but turn out to be very common inside a cloud cluster of an incipient storm (Palmen and Newton 1969; Montgomery et al. 2006). Even though the eyewall pattern depends crucially on the boundary at r = 0, all other features such as the stable behaviors of the SC or the inward convergence are still valid. Those familiar with nonlinear systems will quickly realize that the stable SC obtained above is a kind of kink, a stationary soliton of the VME. The kink with the above features owns its existence solely to the boundary topology (Dirichlet type for both u and w at all boundaries except the Neumann type for u at the top and bottom). Hoverer, one should be especially careful that the kink obtained in this study is not a completely physical kink, whose existence is completely due to the topological structure of boundaries. This is because the VME here is integrated by the *numerical method* and the kink may be resulted from the computational procedures instead of from physical processes. The filter applied over the whole domain consumes a large part of energy of the system and the kink should be named "numerical kink (NK)" to reflect this fact. A little further scrutiny will show that Eqs (1) and (2) in fact accept an infinitive number of trivial solitons (given by any constant number, which is some kind of "*vacuum*"). A systematic treatment should re-write Eqs (1) and (2) in terms of a streamfunction (it is possible due to Eq. (2)). The resultant equation can be regarded as a field equation and one needs to find an appropriate Lorentz-invariant Lagrangian and Hamiltonian for the system from which the full properties of the kink can be studied further (Rubakov 2002). However, this theoretical produce will not be pursued here.

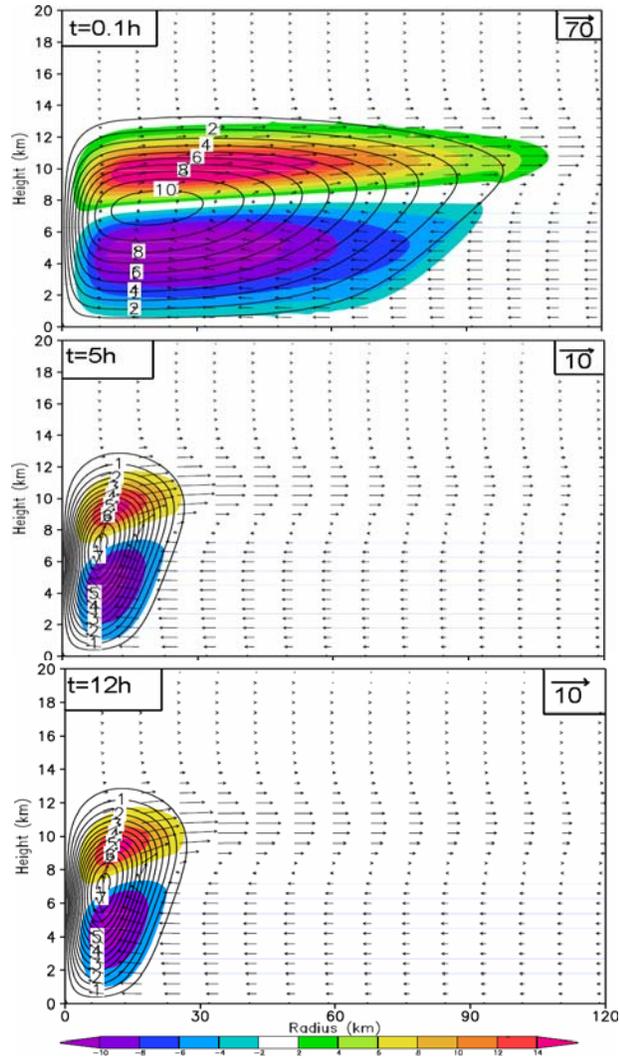

Figure 2. Radius-height cross-sections of the vertical motion (contours) after 0.1-h, 5-h, and 12-h of integration. Superimposed is the in-plane flow at the corresponding time. Shadings denote the total radial and vertical advections from the VME

5. **Sensitivity experiments**

In this section, several sensitivity experiments are conducted to ensure that the results obtained in the control run are not artifacts of numerical procedures. This is a necessary step for any numerical model if one expects to get a sensible solution from the model.

*a. Boundary condition for w at r = 0:* (EXP1)

The results from Section 4 showed an interesting behavior of the SC with the emergence of the eyewall pattern after just few time steps of integration. As mentioned in Section 4, this is due to the boundary for w applied at r = 0. Experiment EXP1 is designed to investigate more about this behavior of the SC. A simple way to do the test is to change the coefficient η in the relationship $w_{r=0}$ =

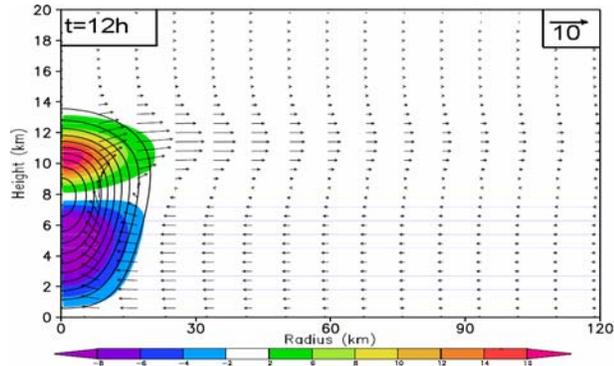

Figure 3. The same as in Fig. 2c but for experiment EXP 1. See text for detail

$\eta w_{r=dr}$. The value $\eta = 0$ is used in the control run and $\eta = 1$ will be used in this experiment (correspond to Neumann boundary $\partial w/\partial r|_{r=0} = 0$). Fig. 3 shows the SC after 12-hour of integration. Even though the SC still reaches the stable configuration after few hours of integration, one no longer sees the eyewall pattern of the SC as in the control run. All other main features such as the shrinkage of the initial cluster into a more concentric vortex at the later time are still valid. The boundary condition for w at r = 0 is therefore essential in deciding the formation of the eye, which is a universal property of a typical kink.

*b. Boundary condition for w at r = R:* (EXP2)

The boundary condition of w at the outer boundary r = R is changed from Neumann to Dirichlet boundary $w|_{r=R} = 0$ to see whether the vertical motion far from the center can have any impact on the stability or structure of the SC in the central region. The results show that the outer boundary virtually has no influences at all to the final solutions (not shown). There is no experiment on the boundaries for both w and u at the top and bottom of the model because of the assumption of the rigidity of the boundaries together with no frictional effects.

*c. Ensemble of vortical hot towers (VHT):* (EXP3)

This experiment utilizes a more realistic initialization profile for vertical motion than that used in the control run. The cloud cluster will

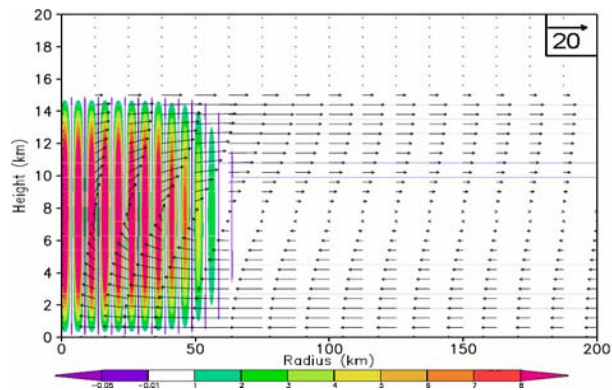

Figure 4. The same as in Fig. 1 but for experiment EXP 3 with an initialization from an ensemble of VHTs

now consist of an ensemble of small-scale deep cumulonimbus convective columns given by the expression:

$$w(r,z,0) = w_0 \left( \sin(\frac{\pi.r}{L}) + 0.9 \right) e^{-(r-a)^2/R_0^2} \sin(\frac{\pi z}{H}) - \varepsilon \qquad (5)$$

where L is the scale of convective columns which is equal to 5 km in this experiment. The additive number 0.9 is added to ensure the existence of the downdrafts in between the VHTs. The incipient cloud cluster associated with a real MCS does not consist simply of continuous updrafts distributed evenly over the whole area as in the control run but contains many small-scale VHTs. This is often observed during the genesis of hurricanes as well as in high-resolution numerical simulations. Fig. 4 shows the initial condition for vertical and radial winds initialized by profile (5). Unlike the control run, the initial radial flow in this experiment has not only a right structure but also more realistic magnitude at the surface. The model integration after 12-h shows similar behaviors of the SC as in the control run: the kink is conserved. Other conclusions regarding the convergence and eyewall structure are still applied.

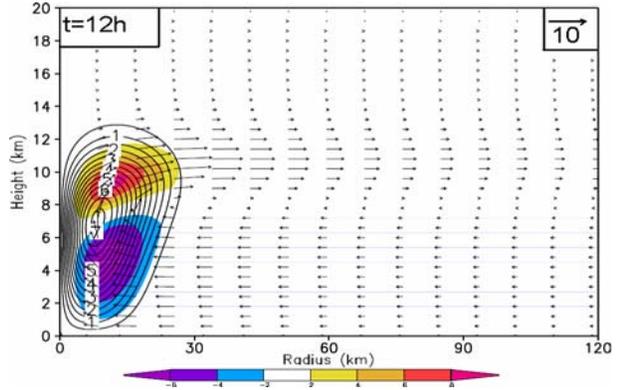

Figure 5. The same as Fig. 2 but for experiment EXP 3 valid at t = 12h after integration

d. *Different stratification of the atmosphere* (EXP4)

In this experiment, the stratification is modified to see how it effects the SC. The stratification at the upper level is kept nearly the same as in the control run but that at the lower levels is assumed a more stable than that in the control run. It turns out that the SC is damped out quickly after few hours of integration (not shown) and no SC is seen after 12-h. This is because too stable atmosphere will convert essentially almost all of the vertical motions into waves over the whole domain, which then propagate quickly away from the center and no stationary configuration is resulted. Even though this sounds physically good, whether it is the case in a real atmosphere is still an open question and needs more investigations. In an additional experiment, the atmosphere is set to be very stable above z = 10 km, and virtually neutral at below (given by the function: $N^2 = \chi \exp(z/H)$ for z > 10 km, and reduce by a factor of two at levels below 10km). Results show that there is no essential change in the behaviors of the SC except that the SC is no longer confined within the troposphere (below 15 km) but expand over the whole domain up to z = 18 km (not shown). Also, the maximal radial convergence occurs at the mid-level rather than at the top of planetary boundary as in the control run.

*e. Different feedback mechanisms:* (EXP5)

Because of the important roles of the feedback assumption, it is of interest to investigate how the results obtained from the control run are altered under different feedback mechanisms. In this experiment, the feedback coefficient (the k parameter in the relation J = kw) will take two different values: $k_1 = 2k$ and $k_2 = 0.5k$ where k is the default value in the control run given in table 1. The simulation with $k_1$ blows up after few time minutes of integration (not shown). This is hardly surprising because, numerically, the strong feedback will result in fast-growing gravity waves that the model can not handle accordingly, a common property of the nonlinear systems. The wave energy is so large that the filter becomes useless. Physically, this is corresponding to the case where the vertical motion is accelerated so enormously that the flow will grow infinitely with time. For the case with coefficient $k_2$, one can anticipate a simple outcome: an unavoidable death for the SC (not shown). The waves consume a larger amount of the total energy than the feedback contribution can compensate. This experiment indicates that hurricanes can only develop when the "right" feedback mechanism is activated. Too weak or too strong both result in unphysical solutions. This explains for the scarcity of hurricanes in nature. One may want to try quadratic or cubic feedback relations but these experiments are not provided here.

6. **Eye formation and its implications**

A remarkable result from the control run is the emergence of a well-defined eye after several hours of model integration. This eyewall structure does not depend on the feedback mechanism but merely on the boundary condition of w imposed at r = 0 as seen in EXP1. Changing the boundary of w at r = 0 results in a complete disappearance of the eye. One may be tempted at this stage to conclude that the eye-formation result therefore may not have any practical implications at all. For a typical hurricane, the appearance of an eye usually signifies the mature stage of the hurricane during which the intensity fluctuation is minimal, and the hurricane can maintain its destructive status for a period of time before landfalling. An important question is how the eye is formed during the transition from the tropical depression to the mature stage. There are two potential ways to imagine the eye formation:

a. If one assumes a bell (Gaussian) shape for vertical motion with the maximal updraft at the center of a cloud cluster at the initial time, then the transition from the bell shape to the M shape may reflect an important characteristic of a nonlinear system:

the transition is a metamorphosis from one unstable solution to another more stable one (Here the author borrows the M-shape to imagine the pattern of vertical motion at the mature stage during which the peaked updrafts are not exactly at the center of the storm but form an annulus around the center with virtually no vertical motion at the central region). This transition may be smooth or it could be an abrupt change. Any small trigger of the environment is able to activate such a transition. In order to fully understand this transition, one has to obtain the full picture of the solutions of the primitive equations as well as their stability behavior. So far, no such solutions exist.

b.  The appearance of an eye is a matter of "*a smart choice*". Observations show that during the genesis of a hurricane, there always exists a broad cloud cluster associated with some MCSs with strong pulses of vertical motion inside the cluster. High-resolution numerical simulations also reveal the existence of many VHTs embedded within the cluster during the genesis phase. These VHTs are characterized by strong updrafts inside and weak descending motions outside their cores (see a sketch in Fig. 6). In the author's opinion, these downdrafts, which have not received enough attentions in previous studies, provide potential embryos for the formation of the eye. The cluster can always pick up a descending area somewhere around the central region and build up the eye around it. It is these downdrafts that give rise to the boundary condition for w at r = 0 as used in the control run. The formation of the eye is virtually always guaranteed because of the availability of the descending areas. However, whether the eye can survive depends on many other factors such as when the feedback is activated or the stratification of the lower levels is too stable or not. The possibility for a cluster to find such a descending location at the right moment with the right feedback will mark a significant change in its lifecycle. As soon as the cluster finds an eye, no matter how small it is, it will grow quickly and build up a concentric eye structure. Unlike the first imagination above, the appearance of eye here is a game of selection rather than an internal behavior of the solution.

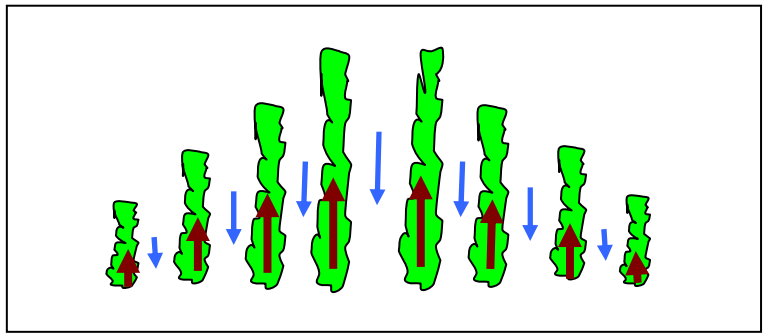

Figure 6. Sketch of the distribution of VHTs within the initial cloud cluster during the genesis of a typical hurricane. The red/blue arrows represent the updrafts/downdrafts

7. **Discussions and conclusions**

Contemporary works on hurricane theory are still limited in approximating the SC as a second-order perturbation compared with the PC. In this way, the SC owns its existence to the PC instead of having its active contributions in the whole development of hurricanes. This classical point of view is dominant in most of the theoretical works of hurricanes up to now. The author's point of view, based on the work of K04, however inclines more on the decisive roles of the SC in establishing the structure and temporal evolution of hurricanes. In this work, the SC under the same feedback mechanism as employed in K04 is investigated, using a numerical method. With the feedback assumption, the VME and the continuity equation form a close system which is enough to solve for the SC completely. The numerical solutions reveal that:
  i. There is no infinite growth of the SC with time
  ii. The radial convergence toward the center is an internal property of the SC
  iii. The SC can develop and maintain itself through feedback mechanism
  iv. There appears a hurricane eye structure during the development of storms provided that a proper boundary condition is applied
  v. Too stable atmosphere at the lower boundary is inimical to hurricane development

The first feature is what we expect for a sensible theory. In K04, the NLTs in the VME are neglected and this results in an exponential growth of the SC with time. This happens with all current theoretical models which attempt to study the temporal evolution of an incipient storm. The inclusion of the NLTs makes the system of the primitive equations too hard to solve for analytically but turns out to be very important in controlling the overall evolution of the system. A numerical kink emerges with finite energy after few hours integrated. The second feature is an interesting result. It tells us that, once an incipient storm appears, the SC associated with the storm will tend to gather peripheral convection toward the center to feed its development. The formation of new peripheral convections will be entrained toward the center to attain a new stable structure. The growth of the storm will therefore depend remarkably on the size of the initial cloud cluster. The larger the cluster, the quicker the storm can grow. This result also points out that the stable structure of hurricanes should be the one with a very-confined eye. The third feature is an arguing property because it depends crucially on the feedback assumption. Ultimately, the feedback is linked to the moisture supply at the surface, which relates to the PC. Therefore, the closure of Eqs. (1) and (2) is not guaranteed. A question is: Is the SC be able to drive the whole development of the PC, or it is the PC that determines the SC? However, the central point here is that, no matter what is the cause, the results in this work imply that the SC is more important than one previously thought. Through the feedback assumption, it can maintain itself and nurture the PC mutually. The vertical advection from the SC plays a key role in determining the vertical structure and the growth of the PC with time (e.g. the tangential wind close to the center grows much faster than that far from the center. See K04 for more detail). The evolution of a storm is one of the most important and interesting stages during the lifetime of a hurricane. By simply treating the SC as a secondary approximation superimposed on the PC, one has throw away the ability to explain for the development of the storm. Moreover, the structure of storm is unrealistic without the SC. One possible reason for the important roles of the SC lies in its capability to provide a strong acceleration under the impacts of the Coriolis torque. One can imagine a situation in which a completely axisymmetric radial motion converges toward a center. Under the Coriolis force, an anticyclonic torque will intensify the tangential wind persistently. The tangential wind therefore attains larger value as it approaches the center. Because of the centrifugal force, pressure will be lowered consequently, which is somewhat the same as one stirs a cup of water. The water surface will be adjusted by the speed one stirs the cup, an example of the wind field determining the mass field (Gill 1982)

The result regarding the formation of hurricane eyes, as one has seen throughout, is ultimately related to the boundary condition at r = 0. The eyewall structure will be more evident if w is zero not just exactly at r = 0 but over an (small) area (e.g. w = 0 inside a cylinder of radius r < 1 km). If one takes into account a further fact that w is slightly negative to reflect the downdraft, one will have a more realistic structure of a typical eye. So the question is whether this result offers a realistic mechanism for the formation of hurricane eyes or it is just an artifact of numerical method. The author's answer is the former, which implies that the downdrafts within a cloud cluster during the genesis phase of hurricanes are the most essential factor in the formation of hurricane eyes. It should be mentioned again that the assumption of positive feedback is essential for the first and third conclusions but has no significant impacts on the rest.

**Acknowledgement**

This work is supported by the research assistantship and partly by the Vietnam Education Foundation fellowship.**REFERENCES**

Charney, J. G., and A. A. Eliassen, 1964: On the growth of the hurricane depression. *J. Atmos. Sci.,* **21**, 68-75.
Emanuel, K. A., 1986: An air-sea Interaction Theory for Tropical Cyclone. Part I: Steady state maintenance, *J. Atmos. Sci.*, 4**3**, 585-604.
\_\_\_\_, 2003: Tropical cyclones. *Annu. Rev. Earth Planet. Sci.*, **31**, 75-104
Gill, A. E., 1982: *Atmosphere-Ocean Dynamics*. Academic Publisher, 662p.
Kieu, C. Q., 2004a: An Analytical Theory for the Early Stage of the Development of Hurricanes: Part I, http://arxiv.org/abs/physics/040707315